\documentclass[aps,prl,superscriptaddress,showpacs,floatfix,twocolumn]{revtex4}

\usepackage{amsmath}
\usepackage[dvipdfm]{graphicx}

\def\pT{\mbox{$p_{\rm T}$}}

\def\v2{\mbox{$v_2$}}

\def\sqrtsNN{\mbox{$\sqrt{s_{NN}}$}}

\begin{document}

\hyphenation{author another created financial paper re-commend-ed Post-Script}

\title{Source breakup dynamics in Au+Au Collisions at \sqrtsNN=200~GeV via 
three-dimensional two-pion source imaging}

\newcommand{\abilene}{Abilene Christian University, Abilene, TX 79699, USA}
\newcommand{\banaras}{Department of Physics, Banaras Hindu University, Varanasi 221005, India}
\newcommand{\bnl}{Brookhaven National Laboratory, Upton, NY 11973-5000, USA}
\newcommand{\caucr}{University of California - Riverside, Riverside, CA 92521, USA}
\newcommand{\cns}{Center for Nuclear Study, Graduate School of Science, University of Tokyo, 7-3-1 Hongo, Bunkyo, Tokyo 113-0033, Japan}
\newcommand{\colorado}{University of Colorado, Boulder, CO 80309, USA}
\newcommand{\columbia}{Columbia University, New York, NY 10027 and Nevis Laboratories, Irvington, NY 10533, USA}
\newcommand{\dapnia}{Dapnia, CEA Saclay, F-91191, Gif-sur-Yvette, France}
\newcommand{\debrecen}{Debrecen University, H-4010 Debrecen, Egyetem t{\'e}r 1, Hungary}
\newcommand{\elte}{ELTE, E{\"o}tv{\"o}s Lor{\'a}nd University, H - 1117 Budapest, P{\'a}zm{\'a}ny P. s. 1/A, Hungary}
\newcommand{\fsu}{Florida State University, Tallahassee, FL 32306, USA}
\newcommand{\gsu}{Georgia State University, Atlanta, GA 30303, USA}
\newcommand{\hiroshima}{Hiroshima University, Kagamiyama, Higashi-Hiroshima 739-8526, Japan}
\newcommand{\ihepprot}{IHEP Protvino, State Research Center of Russian Federation, Institute for High Energy Physics, Protvino, 142281, Russia}
\newcommand{\illuiuc}{University of Illinois at Urbana-Champaign, Urbana, IL 61801, USA}
\newcommand{\isu}{Iowa State University, Ames, IA 50011, USA}
\newcommand{\jinrdubna}{Joint Institute for Nuclear Research, 141980 Dubna, Moscow Region, Russia}
\newcommand{\kaeri}{KAERI, Cyclotron Application Laboratory, Seoul, Korea}
\newcommand{\kek}{KEK, High Energy Accelerator Research Organization, Tsukuba, Ibaraki 305-0801, Japan}
\newcommand{\kfki}{KFKI Research Institute for Particle and Nuclear Physics of the Hungarian Academy of Sciences (MTA KFKI RMKI), H-1525 Budapest 114, POBox 49, Budapest, Hungary}
\newcommand{\korea}{Korea University, Seoul, 136-701, Korea}
\newcommand{\kurchatov}{Russian Research Center ``Kurchatov Institute", Moscow, Russia}
\newcommand{\kyoto}{Kyoto University, Kyoto 606-8502, Japan}
\newcommand{\labllr}{Laboratoire Leprince-Ringuet, Ecole Polytechnique, CNRS-IN2P3, Route de Saclay, F-91128, Palaiseau, France}
\newcommand{\lawllnl}{Lawrence Livermore National Laboratory, Livermore, CA 94550, USA}
\newcommand{\losalamos}{Los Alamos National Laboratory, Los Alamos, NM 87545, USA}
\newcommand{\lpc}{LPC, Universit{\'e} Blaise Pascal, CNRS-IN2P3, Clermont-Fd, 63177 Aubiere Cedex, France}
\newcommand{\lund}{Department of Physics, Lund University, Box 118, SE-221 00 Lund, Sweden}
\newcommand{\muenster}{Institut f\"ur Kernphysik, University of Muenster, D-48149 Muenster, Germany}
\newcommand{\myongji}{Myongji University, Yongin, Kyonggido 449-728, Korea}
\newcommand{\nagasaki}{Nagasaki Institute of Applied Science, Nagasaki-shi, Nagasaki 851-0193, Japan}
\newcommand{\newmex}{University of New Mexico, Albuquerque, NM 87131, USA }
\newcommand{\nmsu}{New Mexico State University, Las Cruces, NM 88003, USA}
\newcommand{\ornl}{Oak Ridge National Laboratory, Oak Ridge, TN 37831, USA}
\newcommand{\orsay}{IPN-Orsay, Universite Paris Sud, CNRS-IN2P3, BP1, F-91406, Orsay, France}
\newcommand{\pnpi}{PNPI, Petersburg Nuclear Physics Institute, Gatchina, Leningrad region, 188300, Russia}
\newcommand{\riken}{RIKEN, The Institute of Physical and Chemical Research, Wako, Saitama 351-0198, Japan}
\newcommand{\rikjrbrc}{RIKEN BNL Research Center, Brookhaven National Laboratory, Upton, NY 11973-5000, USA}
\newcommand{\rikkyo}{Physics Department, Rikkyo University, 3-34-1 Nishi-Ikebukuro, Toshima, Tokyo 171-8501, Japan}
\newcommand{\saispbstu}{Saint Petersburg State Polytechnic University, St. Petersburg, Russia}
\newcommand{\saopaulo}{Universidade de S{\~a}o Paulo, Instituto de F\'{\i}sica, Caixa Postal 66318, S{\~a}o Paulo CEP05315-970, Brazil}
\newcommand{\seoulnat}{System Electronics Laboratory, Seoul National University, Seoul, Korea}
\newcommand{\stonybrkc}{Chemistry Department, Stony Brook University, Stony Brook, SUNY, NY 11794-3400, USA}
\newcommand{\stonycrkp}{Department of Physics and Astronomy, Stony Brook University, SUNY, Stony Brook, NY 11794, USA}
\newcommand{\subatech}{SUBATECH (Ecole des Mines de Nantes, CNRS-IN2P3, Universit{\'e} de Nantes) BP 20722 - 44307, Nantes, France}
\newcommand{\tenn}{University of Tennessee, Knoxville, TN 37996, USA}
\newcommand{\titech}{Department of Physics, Tokyo Institute of Technology, Oh-okayama, Meguro, Tokyo 152-8551, Japan}
\newcommand{\tsukuba}{Institute of Physics, University of Tsukuba, Tsukuba, Ibaraki 305, Japan}
\newcommand{\vandy}{Vanderbilt University, Nashville, TN 37235, USA}
\newcommand{\waseda}{Waseda University, Advanced Research Institute for Science and Engineering, 17 Kikui-cho, Shinjuku-ku, Tokyo 162-0044, Japan}
\newcommand{\weizmann}{Weizmann Institute, Rehovot 76100, Israel}
\newcommand{\yonsei}{Yonsei University, IPAP, Seoul 120-749, Korea}
\affiliation{\abilene}
\affiliation{\banaras}
\affiliation{\bnl}
\affiliation{\caucr}
\affiliation{\cns}
\affiliation{\colorado}
\affiliation{\columbia}
\affiliation{\dapnia}
\affiliation{\debrecen}
\affiliation{\elte}
\affiliation{\fsu}
\affiliation{\gsu}
\affiliation{\hiroshima}
\affiliation{\ihepprot}
\affiliation{\illuiuc}
\affiliation{\isu}
\affiliation{\jinrdubna}
\affiliation{\kaeri}
\affiliation{\kek}
\affiliation{\kfki}
\affiliation{\korea}
\affiliation{\kurchatov}
\affiliation{\kyoto}
\affiliation{\labllr}
\affiliation{\lawllnl}
\affiliation{\losalamos}
\affiliation{\lpc}
\affiliation{\lund}
\affiliation{\muenster}
\affiliation{\myongji}
\affiliation{\nagasaki}
\affiliation{\newmex}
\affiliation{\nmsu}
\affiliation{\ornl}
\affiliation{\orsay}
\affiliation{\pnpi}
\affiliation{\riken}
\affiliation{\rikjrbrc}
\affiliation{\rikkyo}
\affiliation{\saispbstu}
\affiliation{\saopaulo}
\affiliation{\seoulnat}
\affiliation{\stonybrkc}
\affiliation{\stonycrkp}
\affiliation{\subatech}
\affiliation{\tenn}
\affiliation{\titech}
\affiliation{\tsukuba}
\affiliation{\vandy}
\affiliation{\waseda}
\affiliation{\weizmann}
\affiliation{\yonsei}
\author{S.~Afanasiev}	\affiliation{\jinrdubna}
\author{C.~Aidala}	\affiliation{\columbia}
\author{N.N.~Ajitanand}	\affiliation{\stonybrkc}
\author{Y.~Akiba}	\affiliation{\riken} \affiliation{\rikjrbrc}
\author{J.~Alexander}	\affiliation{\stonybrkc}
\author{A.~Al-Jamel}	\affiliation{\nmsu}
\author{K.~Aoki}	\affiliation{\kyoto} \affiliation{\riken}
\author{L.~Aphecetche}	\affiliation{\subatech}
\author{R.~Armendariz}	\affiliation{\nmsu}
\author{S.H.~Aronson}	\affiliation{\bnl}
% ...
\collaboration{PHENIX Collaboration} \noaffiliation

\date{\today}

\begin{abstract}

A three-dimensional (3D) correlation function obtained from mid-rapidity, low \pT , 
pion pairs in central Au+Au collisions at \sqrtsNN=200~GeV is studied. The 
extracted model-independent source function indicates a long range 
tail in the directions of the pion pair transverse momentum (out) and the beam (long). Model comparisons to these distensions indicate a proper breakup time 
$\tau_0 \sim 9$~fm/c and a mean proper emission duration $\Delta\tau \sim 2$~fm/c, leading to  
sizable emission time differences ($\left\langle |\Delta \text{t}_{\text{LCM}}| \right\rangle\approx 12$~fm/c) 
partly due to resonance decays. They also suggest an outside-in ``burning" of 
the emission source reminiscent of many hydrodynamical models.
\end{abstract}
\maketitle

Collisions between heavy nuclei at ultra-relativistic energies 
produce transient systems with energy densities much greater than that 
required to decompose bulk nuclear matter into quarks and gluons~\cite{qgp05}. Such 
systems were predicted to have long lifetimes if a first order phase 
transition occurred during their formation or decay~\cite{pra84}. 

A number of interferometry studies~\cite{koo77} have been made to search for signals of 
long time delays in emissions from actual reaction sources~\cite{lis05}. For 
a Gaussian source function, assumed in the traditional Hanbury Brown Twiss (HBT) 
methodology, this would be signaled by an increase in the width $R$ of the 
emission source function in the out direction of the Bertsch-Pratt 
coordinate system i.e $R_{\rm out}/R_{\rm side} >> 1$. No such result has 
been found by these HBT studies, and the reported Gaussian source functions 
are spheroidal with $R_{\rm out} \approx R_{\rm side}$ in the 
longitudinally co-moving system
(LCMS)~\cite{lis05}. However, a recent study with a 1D source imaging 
technique~\cite{bro97,bro98,bro01}, has observed a 
long non-Gaussian tail in the radial source function and attributed it to 
possible lifetime effects~\cite{chu05,ppg52}. This suggests that further 
study of the source image may give new insights into the reaction dynamics 
leading to source breakup.
%

Here, we extract and 
perform a detailed study of the 3D two-pion source function
using the technique  
proposed by Danielewicz and Pratt~\cite{dan05,dan06}.  
Namely, the 3D correlation function is first decomposed into 
a basis of Cartesian surface-spherical 
harmonics to extract the coefficients, also called moments, of the expansion. 
In turn, they are then imaged or fitted with a trial function to extract the 3D source 
function, which is then used to probe the emission dynamics of the two-pion 
source produced.

Au+Au data (at $\sqrt{s_{NN}}$=200 GeV) was recorded during 2004 with the 
PHENIX detector~\cite{adc03} at the Relativistic Heavy Ion Collider (RHIC). 
The collision vertex $z$ (along the beam axis) 
was constrained to $|z|~<$~30~cm of the nominal crossing point.
Charged pions were detected in the east and west central arms of PHENIX, each 
of which subtends 90$^{o}$ in azimuth $\phi$, and $\pm 0.35$ units of pseudo-rapidity $\eta$.
Tracking and momentum reconstruction were accomplished  
with the drift chamber and two layers of multi-wire proportional chambers 
with pad readout (PC1 and PC3). Particle momenta were measured with a resolution 
$\sigma_p$/p~=~0.7\%~$\oplus 1.0$\%p~(GeV/c). 

Pion identification was achieved for \pT\ $\lesssim 2.0$~GeV/c and \pT\ $\lesssim 1$~GeV/c in
the time of flight and electromagnetic calorimeter, respectively.  For 
this analysis, mid-rapidity pion pairs were selected with $0.2~<~\pT\ <~0.36$~GeV/c from 
semi-central ($0-20\%$) Au+Au collisions at \sqrtsNN=200~GeV, 
where \pT\ is half the pion pair transverse momentum. 
Track merging and splitting effects were removed by appropriate 
cuts on both the real and mixed pair distributions~\cite{ppg52}. 
Systematic variations of these cuts were explored to obtain systematic 
error estimates; on average, they are well within the statistical uncertainty.
Hence, the pair cuts do not introduce any significant bias in the correlation function.
 
The 3D correlation function $C(\mathbf{q}) = N_{\rm fgd}(\mathbf{q})/N_{\rm bkg}(\mathbf{q})$
was constructed as a ratio of 3D relative momentum distribution for 
$\pi^+\pi^+$ and $\pi^-\pi^-$ pairs in the same event $N_{\rm fgd}(\mathbf{q})$ to 
that from mixed events $N_{\rm bkg}(\mathbf{q})$. $C(\mathbf{q})$ is normalised 
to unity for $50<|\mathbf q|<100$~MeV/$c$.
Here, $\mathbf{q}=\frac{(\mathbf{p_1}-\mathbf{p_2})}{2}$ where $\mathbf{p_1}$ 
and $\mathbf{p_2}$ are the momentum 4-vectors in the 
pair center of mass system (PCMS). 
The Lorentz transformation of $\mathbf{q}$ from the laboratory frame to the PCMS is made by a transformation to the pair LCMS along the beam
direction followed by a transformation to the PCMS along the pair transverse momentum~\cite{led04}.

To obtain the moments, the 3D correlation function $C(\mathbf{q})$, is expanded in a 
Cartesian surface-spherical harmonic basis~\cite{dan05,dan06}
\begin{equation}
C(\mathbf{q}) - 1 = R(\mathbf{q}) = \sum_l \sum_{\alpha_1 \ldots \alpha_l}
   R^l_{\alpha_1 \ldots \alpha_l}(q) \,A^l_{\alpha_1 \ldots \alpha_l} (\Omega_\mathbf{q}),
\label{eqn1}
\end{equation}
where $l=0,1,2,\ldots$, $\alpha_i=x, y \mbox{ or } z$, 
$A^l_{\alpha_1 \ldots \alpha_l}(\Omega_\mathbf{q})$
are Cartesian harmonic basis elements; ($\Omega_\mathbf{q}$ is the solid 
angle in $\mathbf{q}$ space); $R^l_{\alpha_1 \ldots \alpha_l}(q)$ are 
Cartesian correlation moments given by Eq.~(\ref{eqn2}); 
and $q$ is the modulus of $\mathbf q$.
\begin{equation}
 R^l_{\alpha_1 \ldots \alpha_l}(q) = \frac{(2l+1)!!}{l!}
 \int \frac{d \Omega_\mathbf{q}}{4\pi} A^l_{\alpha_1 \ldots \alpha_l} 
 (\Omega_\mathbf{q}) \, R(\mathbf{q}).
 \label{eqn2}
\end{equation}
Here, the coordinate axes are oriented so that $z$ (long) is parallel to the beam 
direction, $x$ (out) points in the direction of the total momentum of the pair in 
the LCMS.

Correlation moments, for each order $l$, can be calculated from the 
measured 3D correlation function using Eq.~(\ref{eqn2}). 
Eq.~(\ref{eqn1}) is truncated at $l=6$ and expressed 
in terms of independent even moments only. 
As expected from symmetry considerations, odd moments were found to be 
consistent with zero within statistical uncertainty; 
higher order moments were found to be negligible~\cite{chu07}. 
Up to order 6, there are 10 independent moments: 
$R^0$, $R^2_{x2}$, $R^2_{y2}$, $R^4_{x4}$, $R^4_{y4}$, $R^4_{x2y2}$, 
$R^6_{x6}$, $R^6_{y6}$, $R^6_{x4y2}$ and $R^6_{x2y4}$ 
where $R^2_{x2}$ is shorthand for $R^2_{xx}$ etc (the dependent moments are 
obtained from the independent ones~\cite{dan05,dan06}). These independent moments
were extracted as a function of $q$, by fitting the truncated series to
the measured 3D correlation function with the moments as the parameters of 
the fit. 

\begin{figure}[tbh]
\includegraphics[width=1.0\linewidth]{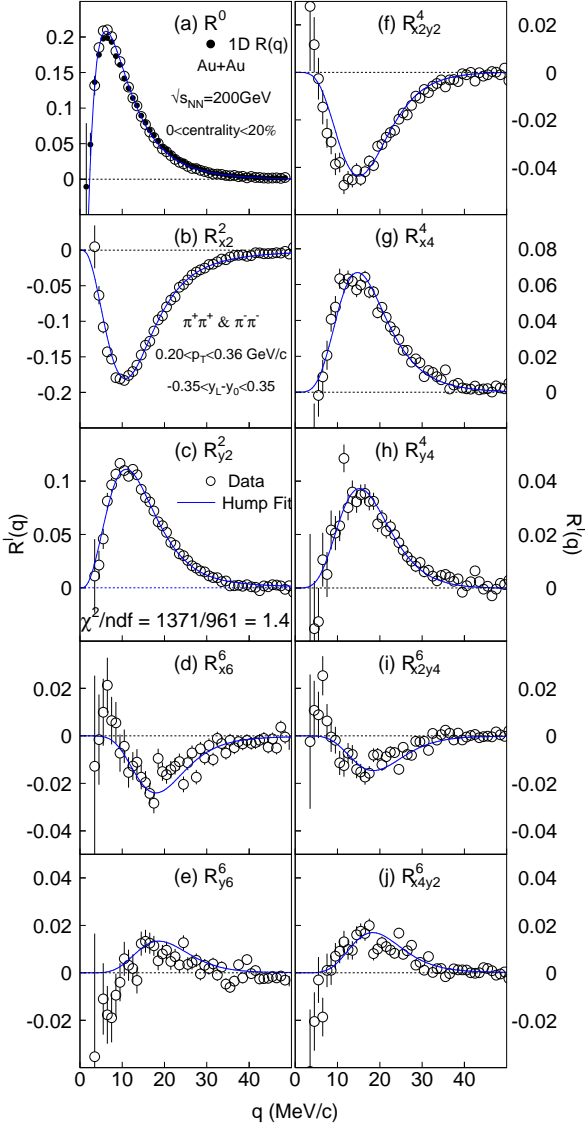}
\vskip -1.cm  
 \caption{\label{phnx_fig1_ppg76}
{Experimental correlation moments $R^l(q)$ for $l=$0, 2, 4, 6. Panel (a) also 
shows a comparison between $R^0(q)$ and $R(q)$. Systematic errors are 
less than the statistical errors. The solid lines indicate
the fit with the Hump function Eq.~(\ref{hump_eqn}).}
}
\end{figure}

Figure~\ref{phnx_fig1_ppg76} shows the correlation 
moments $R^l_{\alpha_1 \ldots \alpha_l}$ 
up to order $l=6$. In panel (a), $R^0(q)$ is shown along with $R(q) = C(q)-1$; 
both represent angle-averaged correlation functions, but they are obtained by very 
different methods. For $R^0(q)$ one uses Eq.~(\ref{eqn2}) and the 3D correlation 
function. For $R(q)$, one simply evaluates the 1D correlation function 
directly as in Ref.~\cite{ppg52}. 

The very good agreement between $R^0(q)$ and $R(q)$ underlines the absence of any 
significant angular acceptance issues and attests to the reliability of the moment 
extraction technique used. 
Figures~\ref{phnx_fig1_ppg76}(b)-(j) show results for the 
moments $l$= 2, 4 and 6. 
Contributions decrease with increasing $l$ in each direction and are relatively 
small for $l= $6. This justifies the truncation of the 
series Eq.~(\ref{eqn1}) at $l= $6. 
The 3D source function $S(\mathbf{r})$ is obtained from these moments 
via imaging or fitting as discussed below.

	Analogous to Eq.~(\ref{eqn1}), $S(\mathbf{r})$ can 
be expanded in a Cartesian Surface-spherical harmonic basis (Eq.~(\ref{eqn3}))

\begin{equation}
 S(\mathbf r) = \sum_l \sum_{\alpha_1 \ldots \alpha_l}
   S^l_{\alpha_1 \ldots \alpha_l}(r) \,A^l_{\alpha_1 \ldots \alpha_l} (\Omega_\mathbf{r}).
\label{eqn3}
\end{equation}
Substitution of the 
series for $R(\mathbf{q})$ and $S(\mathbf{r})$ into the 3D Koonin-Pratt equation~\cite{koo77};
%
\begin{equation}
  C(\mathbf{q})-1 = R(\mathbf{q}) = \int d\mathbf{r} K(\mathbf{q},\mathbf{r}) S(\mathbf{r}),
  \label{3dkpeqn}
\end{equation}
gives a set of 1D relations (Eq.~(\ref{momkpeqn}))~\cite{dan05,dan06}
\begin{equation}
  R^l_{\alpha_1 \ldots \alpha_l}(q) = 4\pi\int dr r^2 K_l(q,r) 
  S^l_{\alpha_1 \ldots \alpha_l}(r),
  \label{momkpeqn}
\end{equation}
which connects the correlation moments $R^l_{\alpha_1 \ldots \alpha_l}(q)$ and source 
moments $S^l_{\alpha_1 \ldots \alpha_l}(r)$. $S(\mathbf{r})$ gives the probability of emitting 
a pair of particles with a separation vector $\mathbf{r}$ in the PCMS. The 3D Kernel, 
$K(\mathbf{q},\mathbf{r})$, incorporates Coulomb interaction and Bose-Einstein symmetrization. 
Strong interaction is assumed to be negligible for pions.
The 1D imaging code of Brown and Danielewicz~\cite{bro97,bro98,bro01}
was used to numerically invert each correlation moment 
$R^l_{\alpha_1 \ldots \alpha_l}(q)$ to extract the corresponding source 
moment $S^l_{\alpha_1 \ldots \alpha_l}(r)$; the latter were then combined as 
in Eq.~(\ref{eqn3}) to obtain the 
source function. 

The 3D source function can also be extracted by directly fitting 
the 3D correlation function with an assumed functional form for 
$S(\mathbf r)$.  This corresponds to a simultaneous fit of the 
ten independent moments.  
A 4-parameter 3D Gaussian (ellipsoid) fit, using MINUIT minimization, 
gives a poor result ($\chi^2/$ndf=3.7).  
The solid line in Fig.~\ref{phnx_fig1_ppg76} 
shows the result of a fit to the independent 
moments with an empirical Hump function given by
\begin{eqnarray}
  \text{S}^H(r_x,r_y,r_z) = \lambda \exp[-f_{\text{s}}(\frac{x^2}{4 r_{x\text{s}}^2} + 
  \frac{y^2}{4 r_{y\text{s}}^2} + \frac{z^2}{4 r_{zs}^2}) \nonumber \\
  - f_{\text{l}}(\frac{x^2}{4 r_{x\text{l}}^2} + \frac{y^2}{4 r_{y\text{l}}^2} + \frac{z^2}{4 r_{z\text{l}}^2})], 
  \label{hump_eqn}
\end{eqnarray}
where $\lambda, r_0, r_{x\text{s}}, r_{y\text{s}}, r_{z\text{s}}, r_{x\text{l}}, r_{y\text{l}}, r_{z\text{l}}$ 
are fit parameters and $f_{\text{s}} = 1/[1 + (r/r_0)^2]$, $f_{\text{l}} = 1 - f_{\text{s}}$.
This 8-parameter Hump function achieves a better fit
to the data ($\chi^2/$ndf=1.4). 
Smearing the track momenta by 
the measured resolution has a negligible effect on the data points and fits.

\begin{figure}[tbh]
\includegraphics[width=1.0\linewidth]{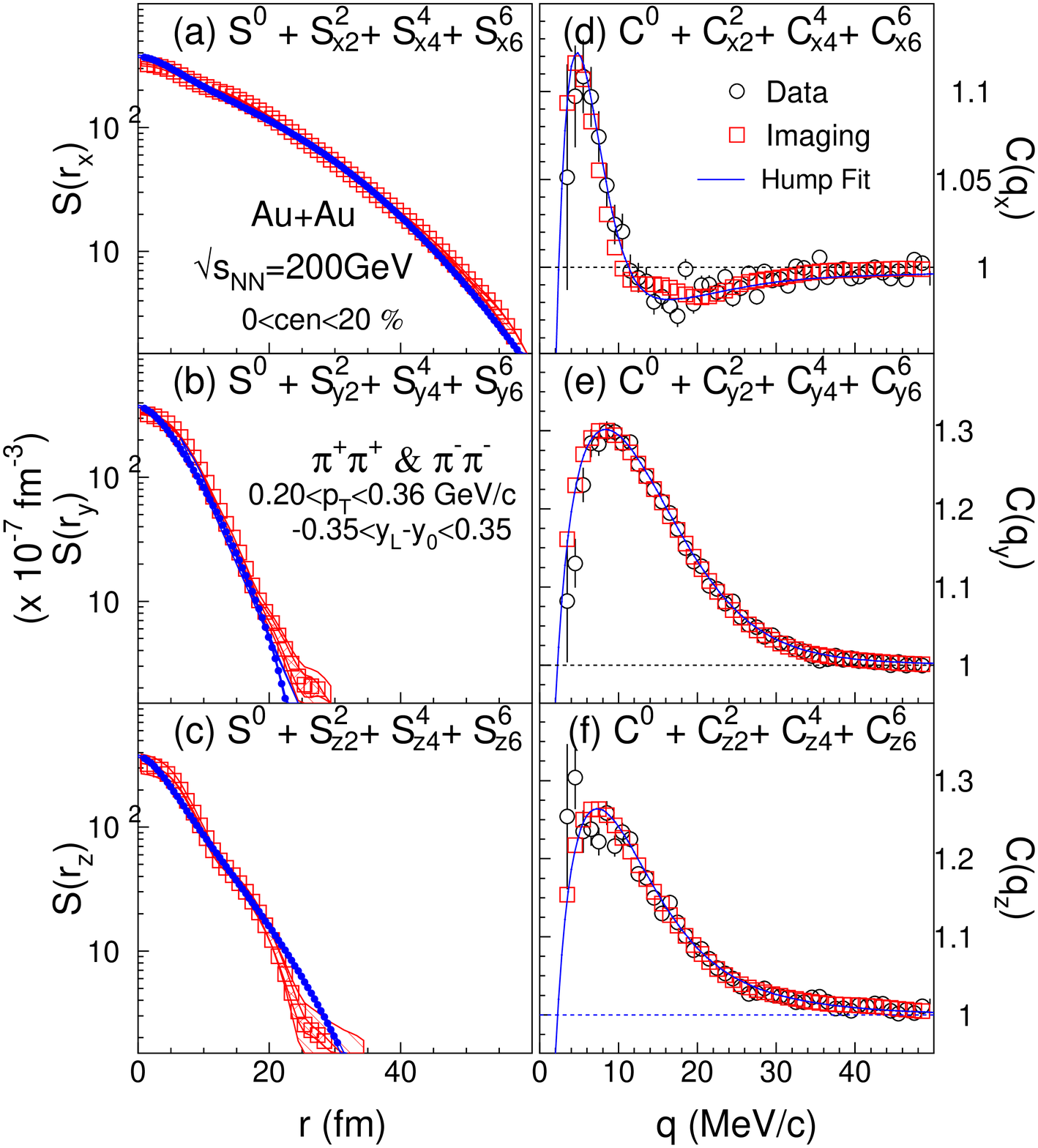}
\vskip -1.cm  
 \caption{\label{phnx_fig2_ppg76}
{Source function profiles S$(r_x)$, S$(r_y)$ and S$(r_z)$ (left panels) and their associated 
correlation profiles C$(q_x)$, C$(q_y)$ and C$(q_z)$ (right panels) in the PCMS. 
Symbols are as indicated. The bands indicate statistical and systematic errors.}
}
\end{figure}

Figure~\ref{phnx_fig2_ppg76}(a)-(c) shows a comparison of profiles of the two-pion 
3D source function in the $x$, $y$ and $z$ directions (S$(r_x)$, S$(r_y)$ and S$(r_z)$)
obtained via fitting (line) and source imaging (squares). 
Source image extraction makes no assumption 
for the shape of the 3D source function; whereas, moment fitting explicitly 
assumes a shape.  Therefore, the good agreement from the two extraction methods
confirms the sufficiency of the Hump function but not its uniqueness.

The function S$(r_x)$ is characterized by a long tail, which is resolved up 
to $\sim$60~fm, in contrast to S$(r_y)$ and S$(r_z)$ which range up 
to $\sim$25~fm. This difference is also reflected in the 
respective correlation profiles (Fig.~\ref{phnx_fig2_ppg76}(d)-(f)) 
obtained by summation of the data (circle), fit (line) and 
image (square) moments up to order $l=6$ (Coulomb effects are not removed).
The broader S$(r_x)$ is associated with the narrower C$(q_x)$ 
(Fig.~\ref{phnx_fig2_ppg76}(a) and (d)), as expected. 

The extended tail lies along the total momentum of the pair in the LCMS. 
Thus, the relative emission times between the pions (including those from 
resonances), as well as the source geometry, will contribute directly 
to S$(r_x)$. The lifetime of the source contributes to the range of 
S$(r_z)$, and S$(r_y)$ reflects its mean transverse 
geometric size. The difference between S$(r_x)$ and 
S$(r_y)$ is thus driven by the combination of the emission time 
difference, the freeze-out dynamics and the kinematic (Lorentz) $\gamma$ boost, 
which is especially important in the out direction.

The event generator Therminator~\cite{kis05,chu07} can shed more light
on the source breakup and emission dynamics. It gives thermal emission from a 
longitudinally oriented cylinder of radius $\rho_{\rm max}$, includes all known resonance 
decays, assumes Bjorken longitudinal boost 
invariance and Blast-Wave transverse expansion with radial velocity $v_r$ semi-linear 
in $\rho$~\cite{kis07}, i.e. $v_r(\rho)=(\rho/\rho_{\rm max})/(\rho/\rho_{\rm max}+ 
v_t)$, where $v_t =1.41$. 
A differential fluid element is a ring defined by cylindrical 
coordinates $z$ and $\rho$; it breaks up at proper time $\tau$ in its rest 
frame or at time $t$ in the lab frame, where $t^2 = \tau^2 + z^2$. 
The freeze-out hypersurface is given by $\tau = \tau_0 + a\rho$, 
where $\tau_0$ is the proper breakup time for $\rho = $0 and $a$ is the space-time 
correlation parameter. In Blast-Wave mode, Therminator sets $a = -0.5$ for source 
emission or ``burning" from outside in as in many hydrodynamical models.

\begin{figure}[tbh]
\includegraphics[width=1.0\linewidth]{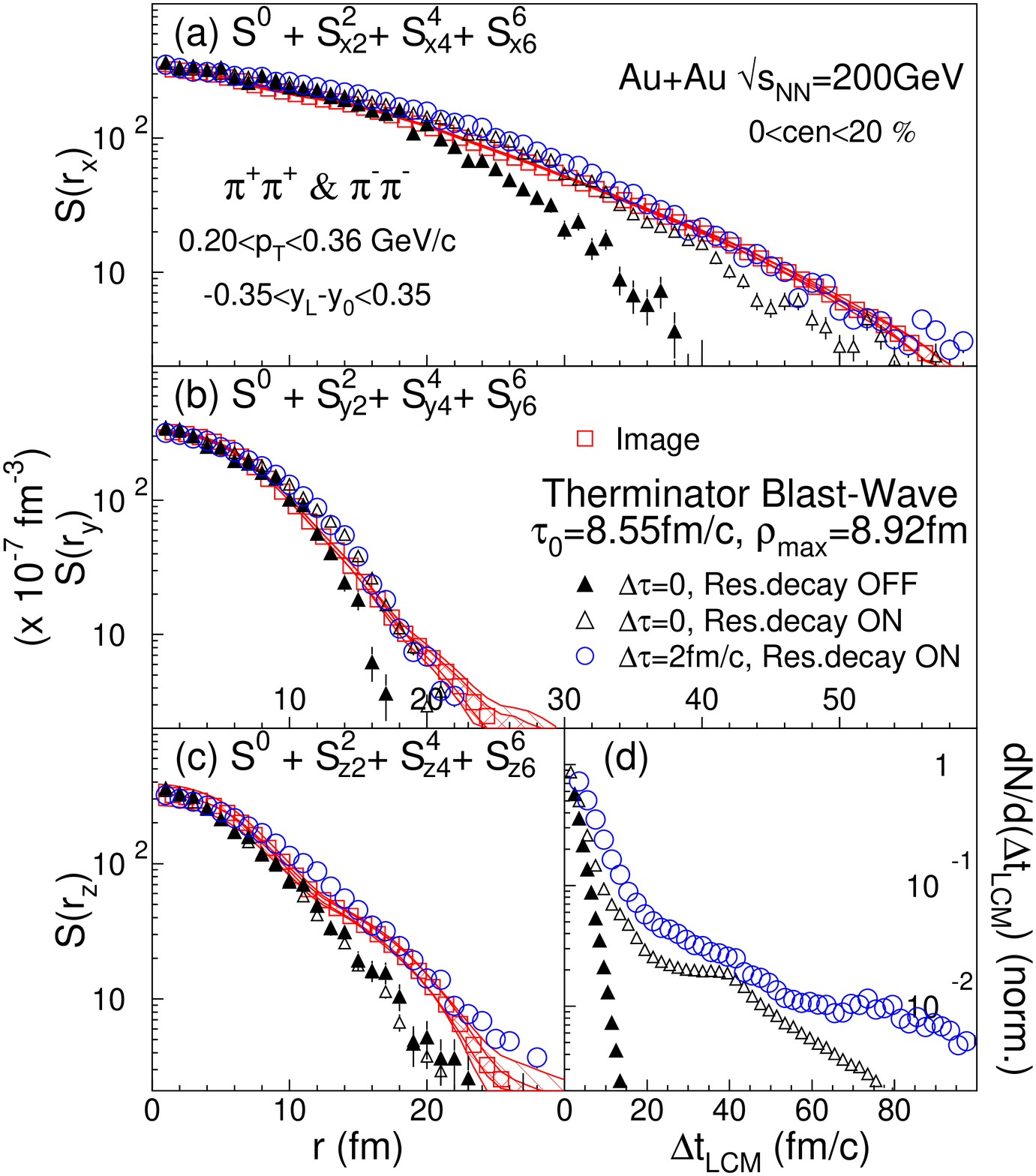}
\vskip -1.cm  
\caption{\label{phnx_fig3_ppg76}
Source function comparison between Therminator calculation and 
image for (a) S$(r_x)$, (b) $S(r_y)$, (c) $S(r_z)$ in PCMS. Panel (d) compares $\Delta \text{t}_{\text{LCM}}$ from Therminator events with various assumptions for $\Delta \tau$ and resonance emission.
}
\end{figure}

Using a set of parameters tuned to fit charged pion and kaon 
spectra~\cite{kis06}, pion pairs from Therminator were obtained with 
the effects of all known resonance decay processes on and off. These pairs were then 
transformed to the PCMS, following the same set of kinematic cuts and Lorentz 
transformations as in the data analysis, to obtain S$(r_i)$ distributions 
for comparison with the data. 

Figure~\ref{phnx_fig3_ppg76} shows that the 3D source function generated by 
Therminator calculations (solid triangles) with $\tau_0=8.55$~fm/c, $\rho_{\rm max}=8.92$~fm 
and other previously tuned parameters~\cite{kis06}, underestimates S$(r_x)$, S($r_y$) and S$(r_z)$. Open triangles (Fig.~\ref{phnx_fig3_ppg76}) show that resonance decays reproduce S($r_y$) ((b)) and extend the 
calculated source function in $x$ ((a))
as expected, but not enough to account for the long tails in S($r_x$) and S($r_z$). 
This suggests that the latter have substantial contribution from pion pairs with 
significantly longer emission time differences. 
Attempts to fit the distributions by only increasing $\tau_0$ or with $a \ge 0$
failed, suggesting a fireball burning from outside in.

The generated distribution of time differences can also be lengthened by 
sampling pions from a family of hypersurfaces defined by a range of 
values of proper breakup times $\tau'$. One such parametrization consists 
of replacing $\tau$ by $\tau'$ chosen from an exponential distribution 
$dN/d\tau' = \frac{\Theta(\tau'-\tau)}{\Delta\tau} \exp[-(\tau'-\tau)/\Delta\tau]$. 
In this parametrization, the width 
of the distribution $\Delta\tau$ represents the mean proper emission duration in 
the rest frame.  Figure~\ref{phnx_fig3_ppg76} shows  that this approach, 
with $\Delta\tau=2$~fm/c (open circles), leads to a fairly good match 
to the observed source profiles in all three directions. A 10\% change in
$\Delta\tau$ spoils this match.

Figure~\ref{phnx_fig3_ppg76}(d) shows the relative emission time distribution in 
the LCMS, $\Delta \text{t}_\text{LCM}$, for pion pairs from events 
with the parameterizations indicated. For a fixed $\tau_0=8.55$~fm/c ($\Delta \tau =0$) and 
resonance decays excluded, the distribution $\Delta \text{t}_\text{LCM}$ is narrow, 
$\left\langle |\Delta \text{t}_{\text{LCM}}| \right\rangle =2.4$~fm/c. The addition of 
resonance decays adds a long tail and gives 
$\left\langle |\Delta \text{t}_{\text{LCM}}|\right\rangle =8.8$~fm/c. Replacing $\tau$ with the 
exponential distribution $\tau'$ with $\Delta \tau=2$~fm/c, results in a $\Delta \text{t}_\text{LCM}$ distribution which is significantly broadened to give 
$\left\langle |\Delta \text{t}_\text{LCM}| \right\rangle =11.8$~fm/c. The wider distribution 
of time delays is needed to reproduce the source distributions. This implies a finite 
non-zero proper emission duration in the emission rest frame. Note that this $\Delta \text{t}_\text{LCM}$ distribution broadening has only a small effect on S($r_y$). 

Figure~\ref{phnx_fig3_ppg76} shows that substantial time 
differences $\Delta \text{t}_\text{LCM}$ are required by the source distensions; however, the interplay 
between proper time and breakup dynamics is model dependent. 
The picture which emerges from the data, in the context of the Therminator model, 
is consistent with that of an expanding fireball with proper breakup 
time $\tau_0 \sim 9$~fm/c, which hadronizes and emits particles over a short but 
non-zero mean proper emission duration $\Delta\tau = 2$~fm/c.

In summary, a new model-independent, three-dimensional source imaging 
technique has been applied to extract the 3D pion emission source 
function in the PCMS frame from Au+Au collisions at $\sqrt{s_{NN}}=200$~GeV. 
The source function has a much greater extent in the 
out ($x$) and long ($z$), than in the side ($y$) direction.   Therminator 
model comparison indicates a fireball burning from outside in with proper lifetime 
$\tau_0 \sim 9$~fm/c and a mean emission duration $\Delta\tau \sim 2$~fm/c, 
leading to  significant relative emission times 
($\left\langle |\Delta \text{t}_\text{LCM}| \right\rangle\approx 12$~fm/c), 
including those due to resonance decay. 



We thank the staff of the Collider-Accelerator and 
Physics Departments at BNL for their vital contributions.  
We thank P. Danielewicz and S. Pratt for their interest and input.
We acknowledge support from 
the Office of Nuclear Physics in DOE Office of Science and NSF (U.S.A.), 
MEXT and JSPS (Japan), 
CNPq and FAPESP (Brazil), 
NSFC (China), 
IN2P3/CNRS, and CEA (France), 
BMBF, DAAD, and AvH (Germany), 
OTKA (Hungary), 
DAE (India), 
ISF (Israel), 
KRF and KOSEF (Korea), 
MES, RAS, and FAAE (Russia),
VR and KAW (Sweden), 
U.S. CRDF for the FSU, 
US-Hungarian NSF-OTKA-MTA, 
and US-Israel BSF.

%

\end{document}